\documentclass[hidelinks,pra,aps,twocolumn]{revtex4}%
\usepackage[utf8]{inputenc}%
\usepackage{amsmath}%
\usepackage{amsfonts}%
\usepackage{amssymb}
\usepackage{tikz}%
\usepackage{graphicx}%
\usepackage{lmodern}%
\usepackage{float}%
\usepackage{bibrevtex4}
\usepackage{wrapfig}
\usepackage{microtype}
\graphicspath{{.}{Graphics/}}
\begin{document}
\title{Realization of Shor's Algorithm at Room Temperature}%
\author{Niklas Johansson}
\email{niklas.johansson@liu.se}
\affiliation{Institutionen för systemteknik, Linköpings Universitet, 581 83
   Linköping, Sweden}

\author{Jan-Åke Larsson}
\email{jan-ake.larsson@liu.se}
\affiliation{Institutionen för systemteknik, Linköpings Universitet, 581 83
    Linköping, Sweden}

\newcommand{\ket}[1]{\ensuremath{\left\vert{#1}\right\rangle}}
\newcommand{\bra}[1]{\ensuremath{\left\langle{#1}\right\vert}}
\newcommand{\alert}[1]{{\color{red}{#1}}}

\begin{abstract}\noindent
	Shor's algorithm can find prime factors of a large number more efficiently 
	than any known classical algorithm. Understanding the properties that gives 
	the speedup is essential for a general and scalable construction. Here we 
	present a realization of Shor's algorithm, that does not need \emph{any} 
	of the simplifications presently needed in current experiments and also 
	gives smaller systematic errors than any former experimental 
	implementation. Our realization is based on classical pass-transistor 
	logic, runs at room temperature, and uses the same amount of resources as a 
	scalable quantum computer. In this paper, the focus is \emph{not} on the 
	result of the factorization, but to	compare our realization with current 
	state-of-the-art experiment, factoring 15. Our result gives further insight 
	to the resources needed for quantum computation, aiming for a true 
	understanding of the subject.
\end{abstract}\thispagestyle{empty}
\maketitle


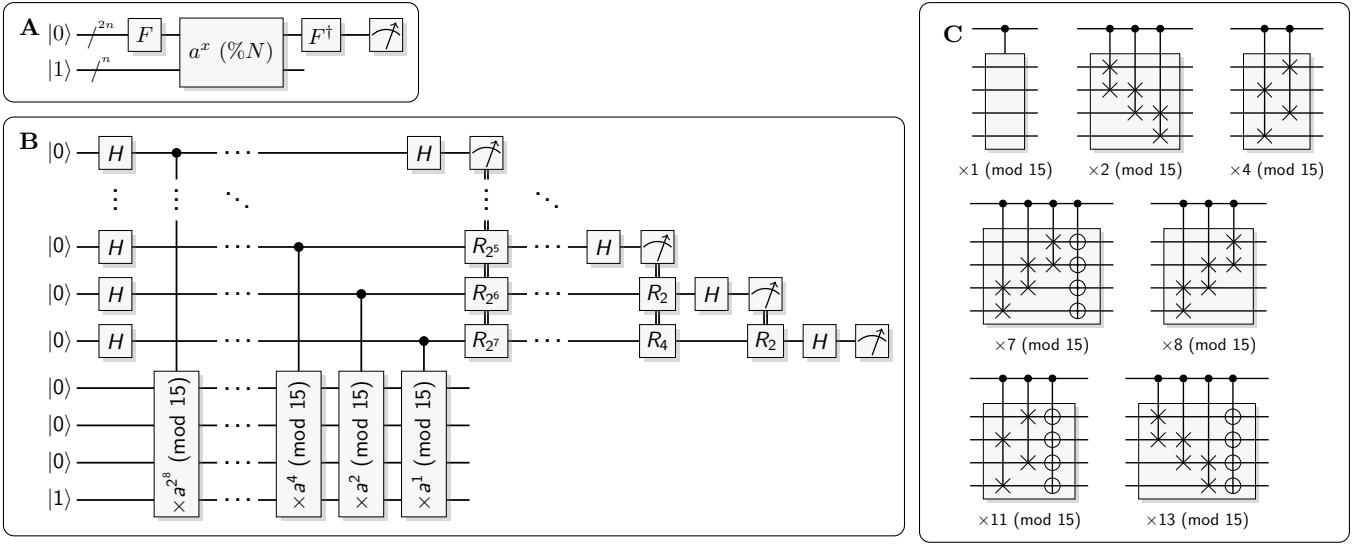
\begin{figure*}[t]
	\begin{center}
	\begin{tikzpicture}[anchor=north west,rounded corners]
	\node [draw,outer sep=1mm](a) {\quad\includegraphics[scale=0.83]{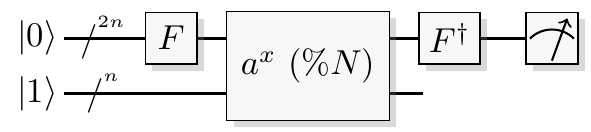}};%
	\node at (a.north west) [anchor=north west,outer sep=2mm] {\bf A};%
	\node at (a.south west) [draw,outer sep=1mm,anchor=north west] (b) 	{\quad\includegraphics[scale=0.83]{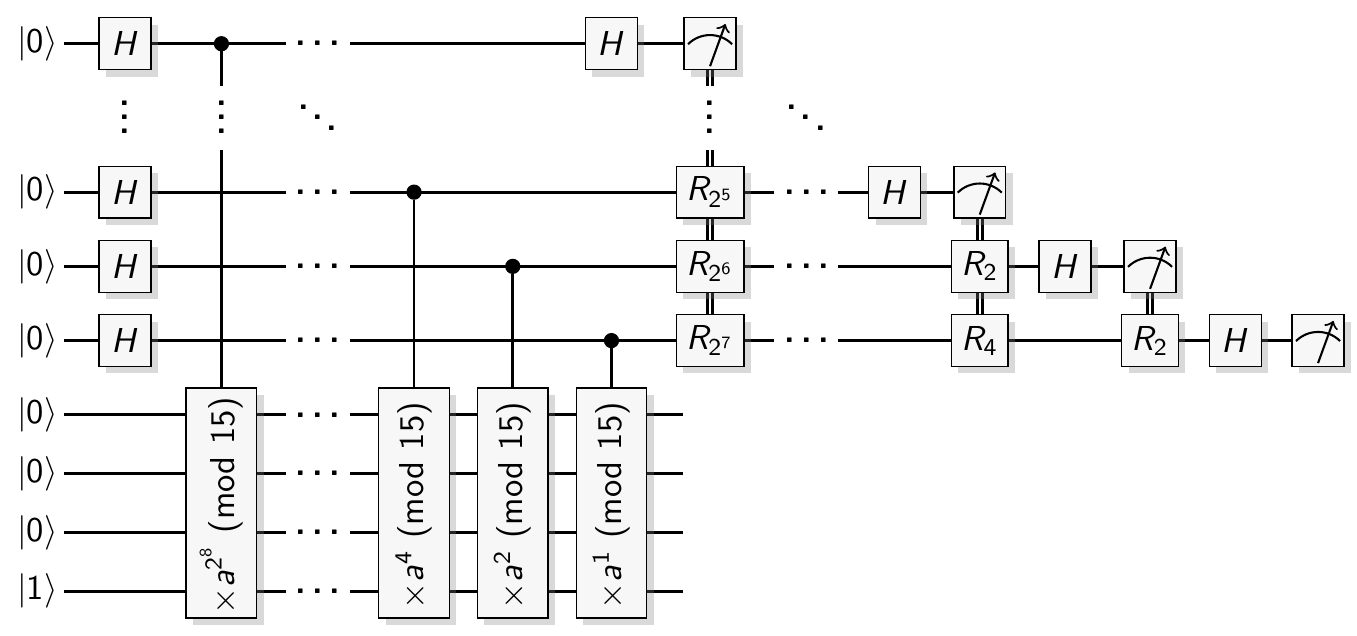}};%
	\node at (b.north west) [anchor=north west,outer sep=2mm] {\bf B};%
	\node at (a.north-|b.north east) (c) [draw,outer sep=1mm,inner sep=2mm] {\quad\tikz{
	\node (1) [inner sep=-1mm]{\includegraphics[scale=.65]{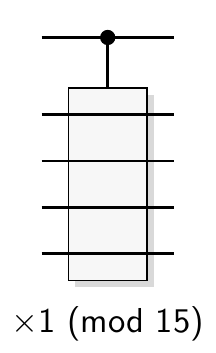}};
	\node at (1.north east) (2)[inner sep=-1mm] {\includegraphics[scale=.65]{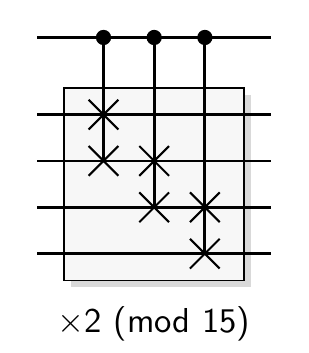}};
	\node at (2.north east) (4)[inner sep=-1mm] {\includegraphics[scale=.65]{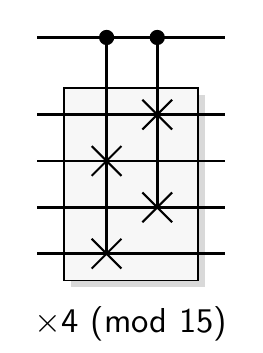}};
	\node at (1.south west) (5)[inner sep=-1mm] {\includegraphics[scale=.65]{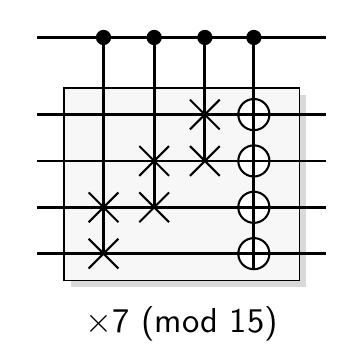}};
	\node at (5.north east) (8)[inner sep=-1mm] {\includegraphics[scale=.65]{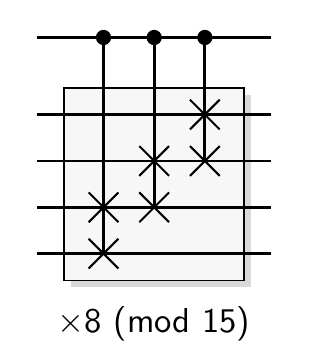}};
	\node at (5.south west) (11)[inner sep=-1mm] {\includegraphics[scale=.65]{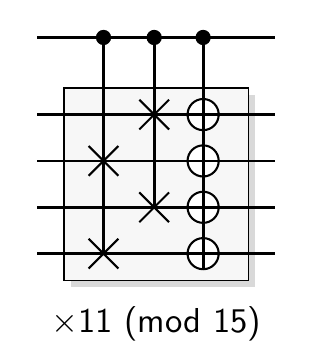}};
	\node at (11.north east) (13)[inner sep=-1mm] {\includegraphics[scale=.65]{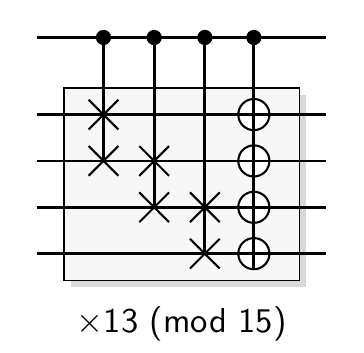}};}};%
	\node at (c.north west) [anchor=north west,outer sep=3mm] {\bf C};%
	\end{tikzpicture}
	\end{center}
	\caption{\textbf{A.} Circuit diagram of the quantum subroutine used in Shor's 
		algorithm. A $2n$-qubit register is initiated in the zero-state $\ket{0}$, 
		and an $n$-qubit register in $\ket{1}$. Basis change of the input-register part 
		of the controlled modular exponentiation operator allow for sampling a 
		probability distribution with peaks at $s/r$.
		\textbf{B.} Shor's algorithm with semiclassical inverse Fourier 
	transform.  Note that 
    $\times2^2\equiv \times7^2 \equiv \times8^2 \equiv \times13^2 
    \equiv \times4^1$ and $\times4^2\equiv\times11^2\equiv\times1$ (mod $15$), 
    so that many of the controlled multiplications will be identities. 
    Therefore most rotations $R_{2^k}$ will never be 
	applied (in the ideal situation), in fact only the very last $R_2$ 
	operation can ever occur. \textbf{C.} 
	Controlled modular multipliers that occur in Shor's algorithm.}
	\label{fig:Shor}
\end{figure*}
\begin{figure*}[t!]
	\begin{center}
		\begin{tikzpicture}[anchor=north west,rounded corners]
		\node [draw,outer sep=1mm,inner sep=-1mm](a) {\quad\includegraphics{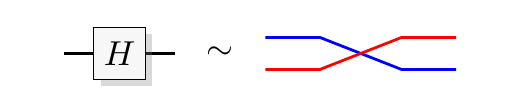}};%
		\node at (a.north west) [outer sep=2mm] {\bf A};%
		\node [draw,outer sep=1mm,inner sep=-1mm] at (a.south west) (b) {\quad\includegraphics{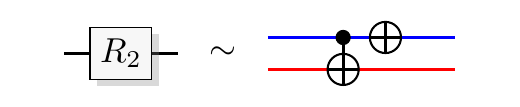}};%
		\node at (b.north west) [outer sep=2mm] {\bf B};%
		\node [draw,outer sep=1mm,inner sep=-1mm] at (a.north east) (c) {\quad\includegraphics{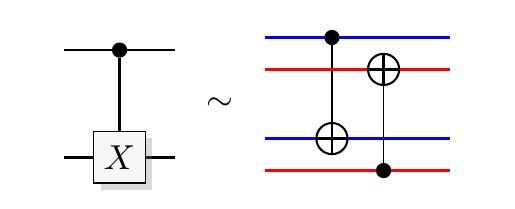}};%
		\node at (c.north west) [outer sep=2mm] {\bf C};%
		\node [draw,outer sep=1mm,inner sep=-1mm,anchor=west] at (c.east) (d) {\quad\includegraphics{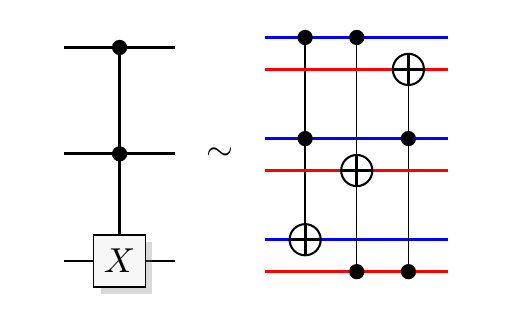}};%
		\node at (d.north west) [outer sep=2mm] {\bf D};%
		\end{tikzpicture}
	\end{center}
	\caption{QSL constructions, computational bit in blue drawn above the phase bit in red,
		\textbf{A.} Hadamard gate, \textbf{B.} Phase gate, \textbf{C.} CNOT gate, \textbf{D.} Toffoli gate.}
	\label{fig:QSL_gates}
\end{figure*}
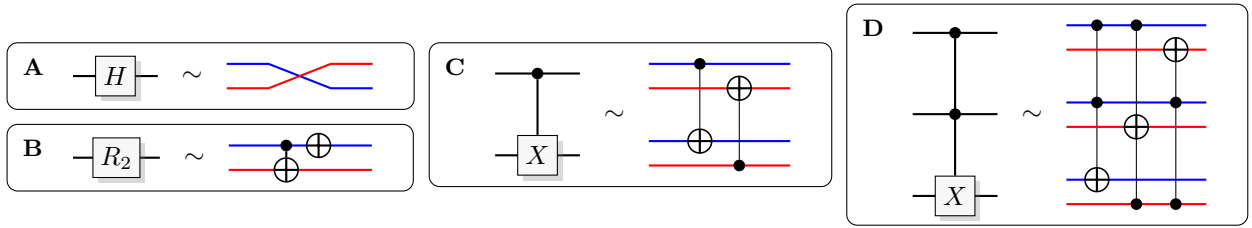

\noindent
Shor's algorithm is one of the few quantum algorithms that solves a
computational problem with real-world applications: to efficiently find a 
factor $q$ in a composite number $N$, which is otherwise thought to be hard. In 
fact, the hardness of finding the integer factorization of a composite number, 
is one of the most widely believed conjectures in computer science, and 
cryptographic applications that we all use in our daily life are built upon 
this.

While large high-fidelity quantum computers are still far away, several
experimental realizations of Shor's algorithm for small numbers have been
presented~\cite{Monz2016,Lucero2012,Martin-Lopez2012,Politi2009,Lu2007,Lanyon2007,Vandersypen2001}.
These are all very impressive demonstration of quantum optimal control, but 
experimental realization of Shor's algorithm with the currently available 
technology is demanding, and this has led to the need for
vast simplifications in the algorithm. There are essentially two parameters 
subject to optimization with regard to circuit size, bit-depth and 
circuit-depth, see \cite{Markov2012} and citations therein. Also the 
approximate quantum Fourier transform \cite{Coppersmith2002} is crucial for 
scalability. In this optimization procedure, one has to
be careful not to over-simplify, or make explicit (or implicit) use of 
knowledge about the solution~\cite{Smolin2013}. 

Shor's algorithm for integer factorization~\cite{Shor1994} finds the order (or
period) of an element $a$ in the multiplicative group of integers modulo $N$. 
Here, the order is the smallest integer $r$ such that $a^r=1$ (mod $N$). This is
sufficient information to find a factor in $N$. The algorithm makes use of an
input-register quantum state, containing an integer $x$, and an output-register 
in which modular exponentiation $a^x$ (mod $N$) is computed. By changing the 
quantum state in the input-register using the quantum Fourier transform, 
performing the calculation, and then inverting the transform, one can with high 
probability retrieve sufficient information to calculate the order, see 
Figure~\ref{fig:Shor}A. More specifically, this procedure let us sample 
from a probability distribution with peaks at $s/r$, where $s$ is uniformly 
distributed over the integers between $0$ and $r-1$.  

Ideally the peaks are completely localized to $s/r$ but in most cases there is 
peak broadening due to Fourier leakage, and to ensure that the measurement 
yields a binary fraction sufficiently close to $s/r$, the input-register needs 
to be at least $2n$ qubits in size, where $n$ is the size of the 
output-register which is large enough to perform calculations mod $N$. 
The full procedure to retrieve $r$ is as follows: 
\begin{enumerate}
	\item Pick at random an integer $a\neq \pm1$ modulo $N$. 
	If GCD$(a,N)$ is a nontrivial factor of $N$, we have a solution.
	\item Otherwise generate, setup, and run the quantum subroutine of 
	Figure~\ref{fig:Shor}A to find a candidate for $s/r$.
	\item Use the continued fraction expansion to retrieve $r$ (or a
	factor in $r$ when $s$ and $r$ has a common factor). 
	\item If $r$ is even, one of GCD$(a^{r/2}\pm 1, N)$ may be a nontrivial 
	factor of $N$. This happens with high probability.
\end{enumerate}

For our example $N=15$ the possible integers that can occur in steps 2-4 are 
$a\in\{2,4,7,8,11,13\}$. One should beware of simplification such as 
``compilation'', where the element $a$ is chosen deliberately to give a short 
period that is easy to find. It is therefore important that the element $a$ is 
chosen randomly~\cite[see e.g.,][]{Smolin2013}. In what follows, we have used 
all alternatives; this is of course only possible because of the small $N$ used.

Some useful simplifications \textit{are} allowed. The Fourier transform in 
Figure~\ref{fig:Shor}A can be exchanged for Hadamard gates, while the inverse Fourier transform can be exchanged for 	Hadamards followed by classically controlled single qubit rotations \cite{Griffiths1996}; by 
advancing the measurement of the controlling qubit and using the outcome as a 
classical control. This decouples the $2n$ qubits of the input-register in the 
sense that the procedure of preparation, transformation, and measurement can be 
performed individually on each qubit. It is common to perform these single 
qubit procedures in sequence on one single qubit, a method known as qubit 
recycling, which reduces the overall bit-depth from $3n$ to $n+1$ at the cost 
of circuit-depth.

\begin{figure*}[t!]
    \begin{center}
        \includegraphics[width=1\linewidth]{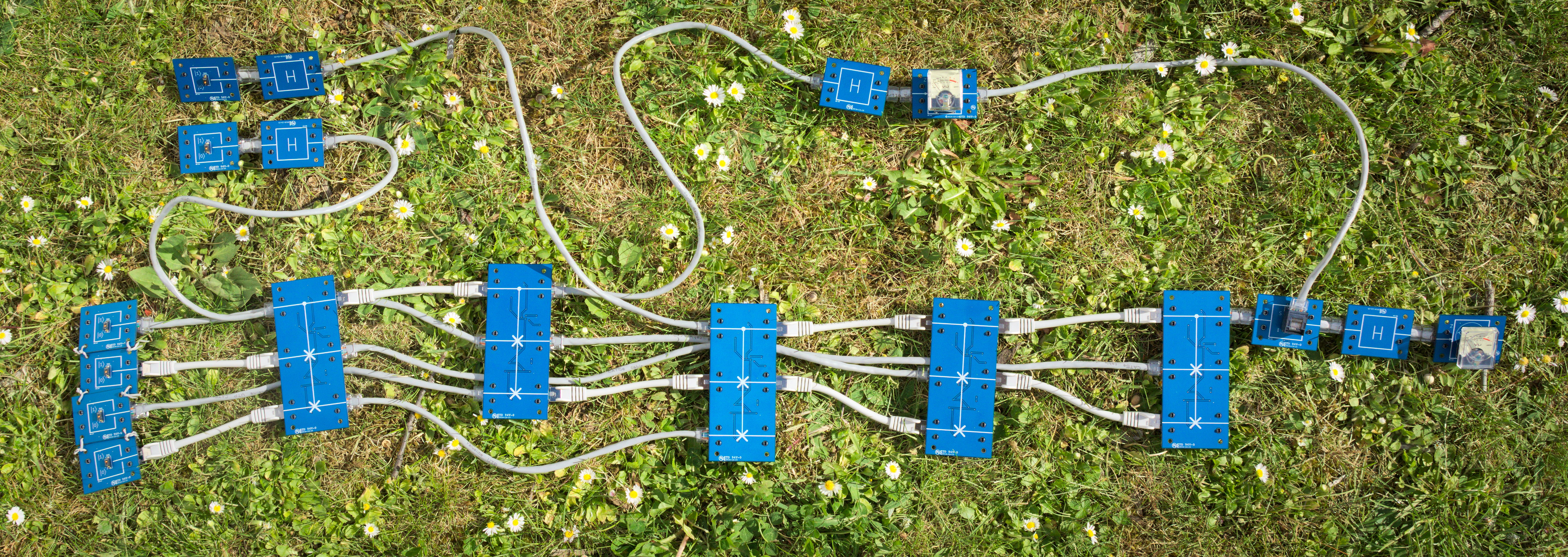}%
    \end{center}
    \caption{QSL realization of Shor's algorithm, when $a=8$ so that the 
    modular multipliers used are, from left to right, $\times8^2\equiv\times4$ 
    (mod 15), and $\times8$ (mod 15).}
    \label{fig:QSL_lawn}
\end{figure*}

\begin{figure*}[t!]
	\begin{center}
		\includegraphics[scale=.39,clip,trim=.5cm 0cm 12.15cm 0cm]{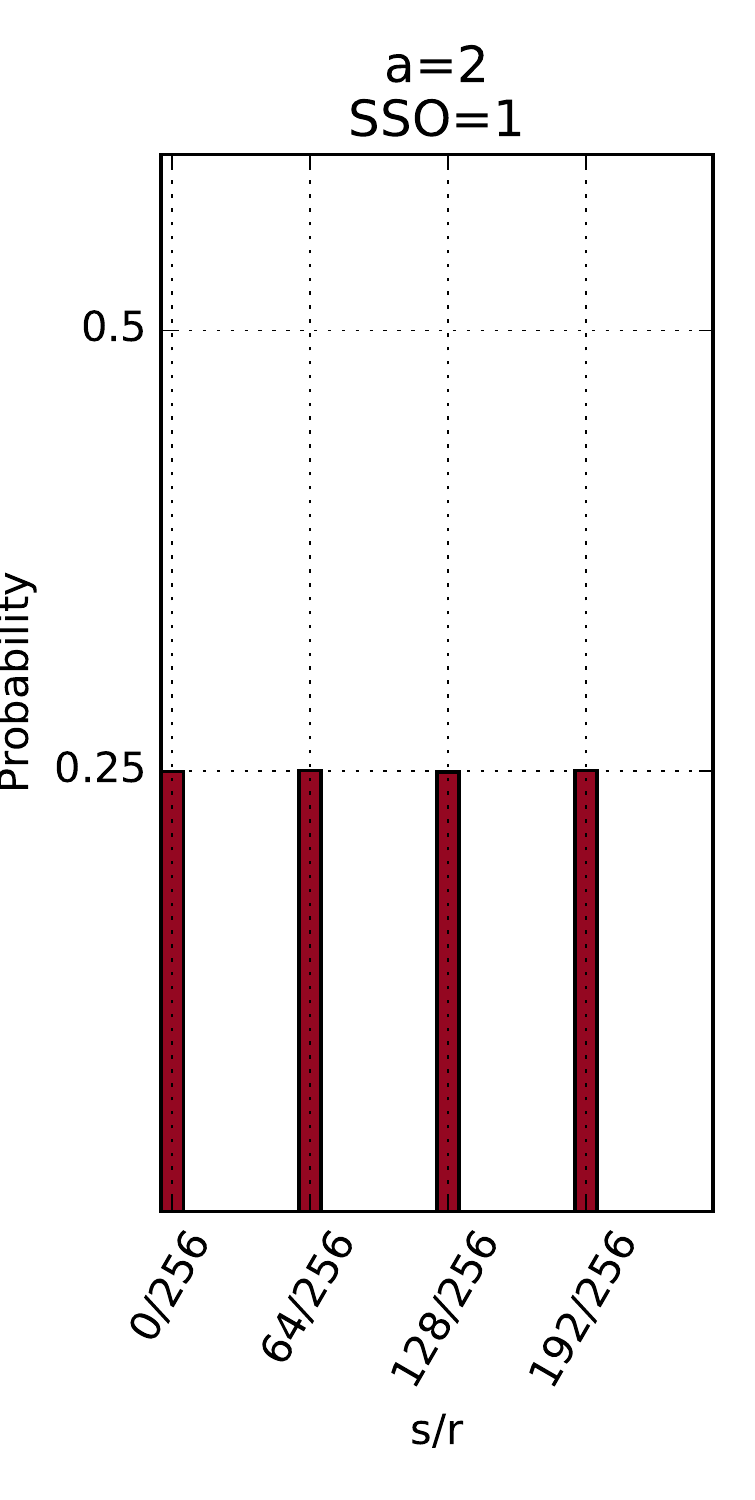}
		\hspace{0cm}\includegraphics[scale=.39,clip,trim=0cm 0cm 12.55cm 
		0cm]{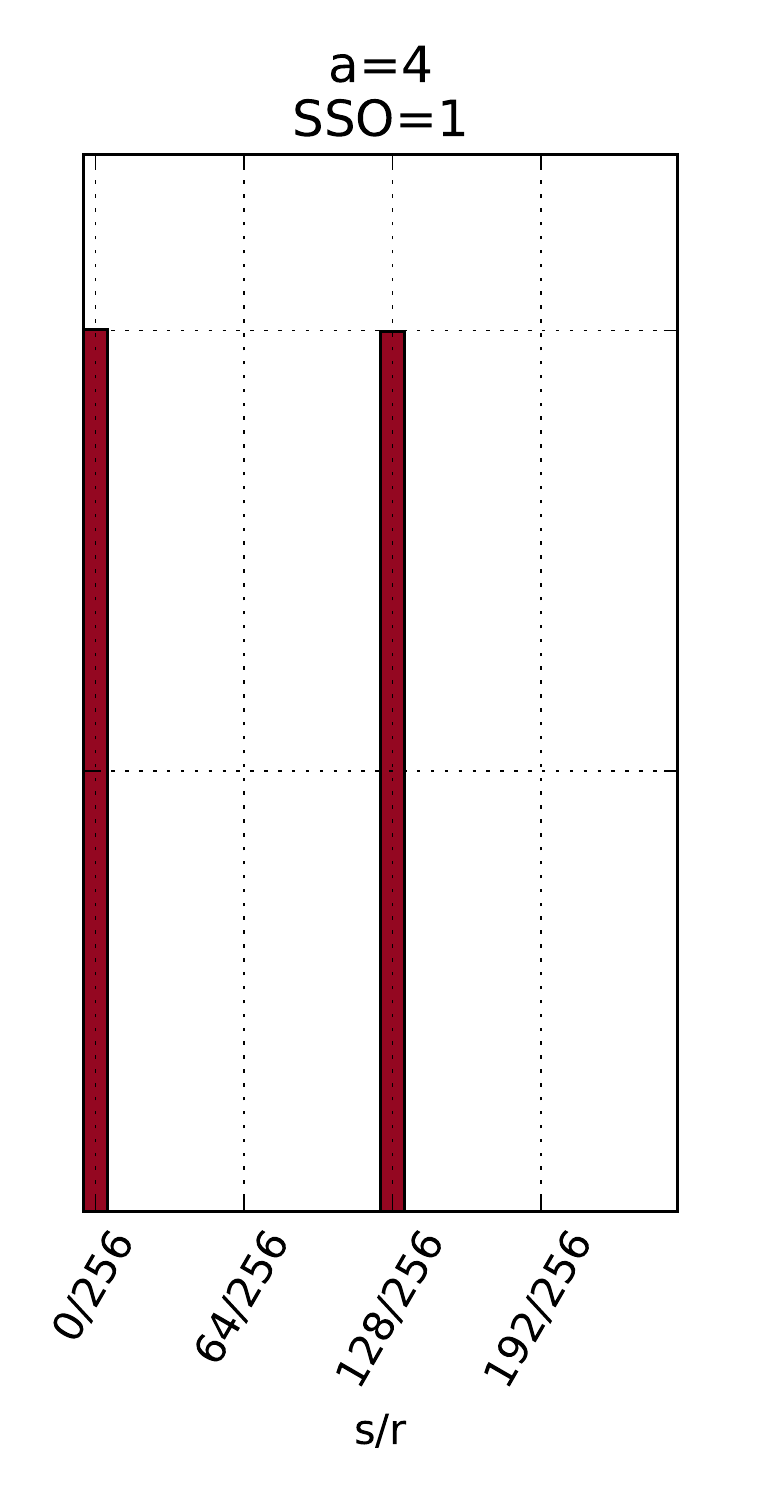}
		\hspace{0cm}\includegraphics[scale=.39,clip,trim=.5cm 0cm 12.55cm 
		0cm]{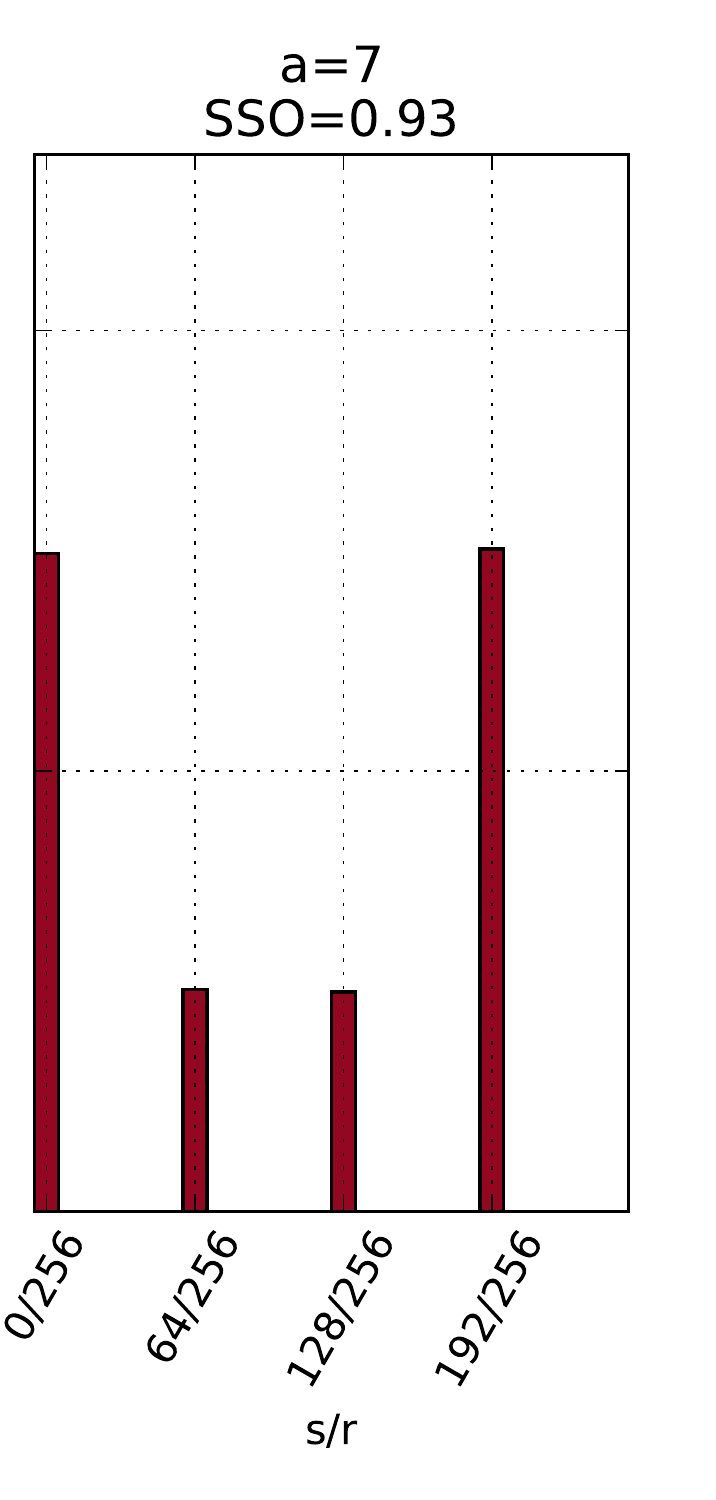}
		\hspace{-0cm}\includegraphics[scale=.39,clip,trim=.5cm 0cm 12.55cm 
		0cm]{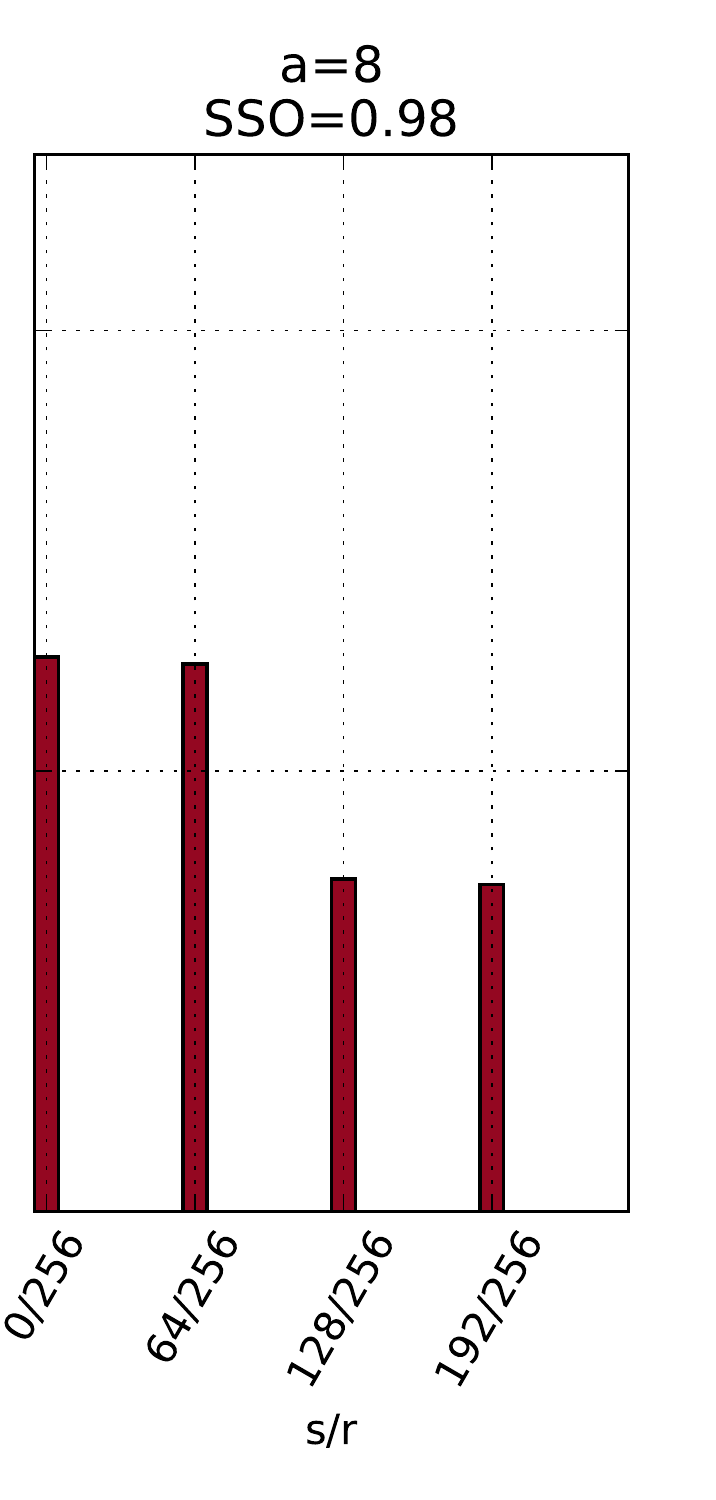}
		\hspace{-0cm}\includegraphics[scale=.39,clip,trim=.5cm 0cm 12.55cm 
		0cm]{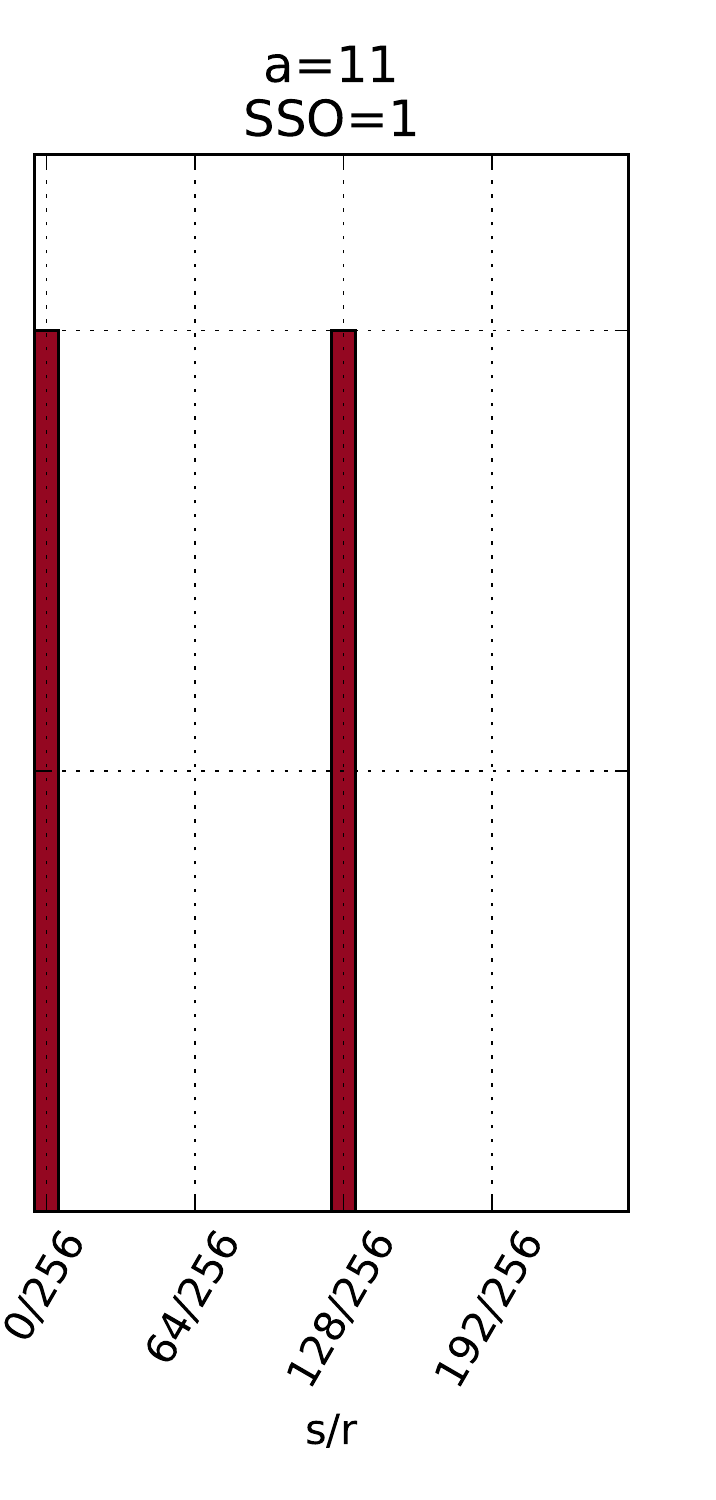}
		\hspace{-0cm}\includegraphics[scale=.39,clip,trim=.5cm 0cm 12.55cm 
		0cm]{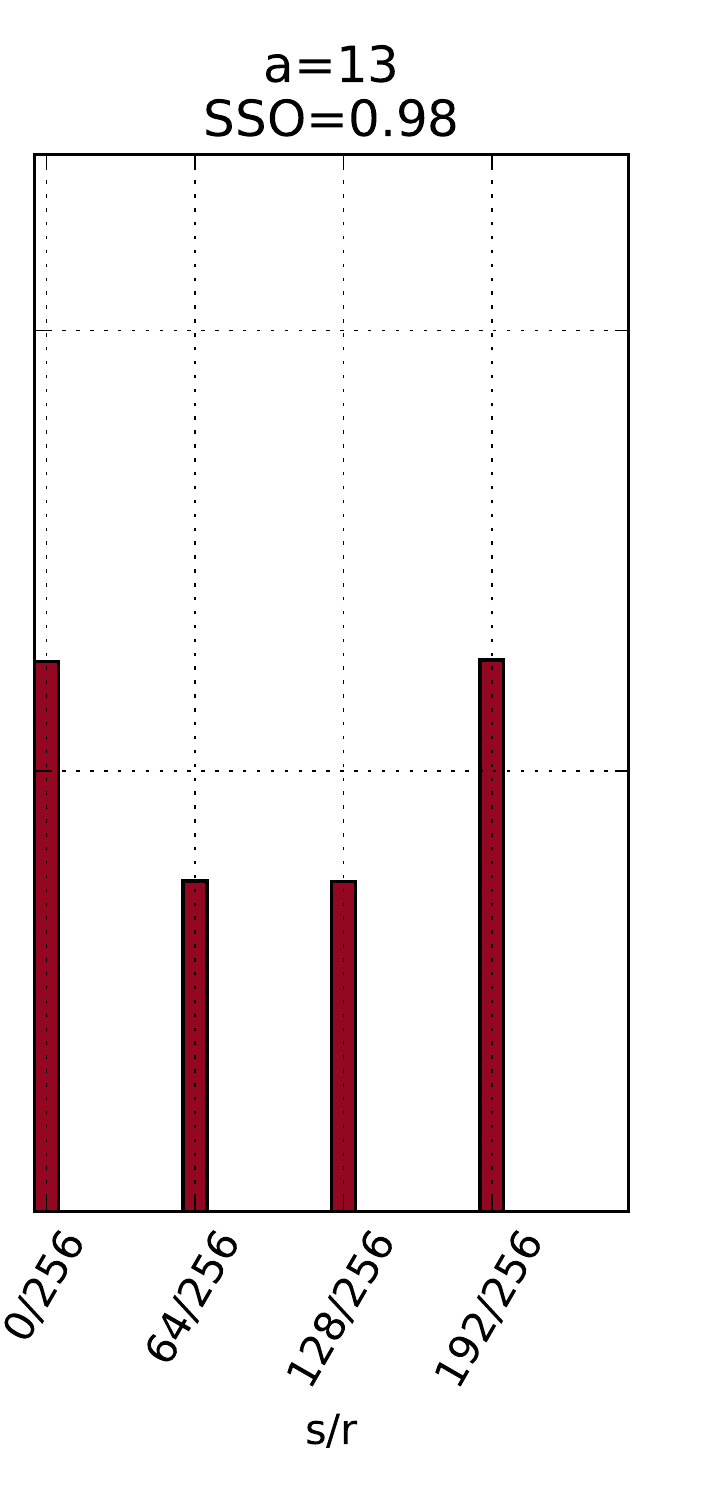}
	\end{center}
	\caption{Estimated output probability distributions of the subroutine 
		for all non-trivial elements $a$ in the multiplicative group of integers mod
		15. Each plot (distribution) is estimated from $10^6$ samples.}
	\label{fig:distribution}
\end{figure*}

The demonstration of Monz \textit{et al}~\cite{Monz2016} is the most advanced 
to date. They use non-Clifford group operations which is absolutely necessary to
demonstrate the advantage of quantum 
computation~\cite{Gottesman1998,Aaronson2004}, and also
refrain from ``compiling'' the circuitry. One simplification they do use is 
restricting the resolution of the input register from $2n$ qubits to three 
qubits. This is possible because all elements in the multiplicative group mod 
15 has power-of-two periods, and Monz \textit{et al} only verify the behavior 
of the exponentiation until and including the first that is equivalent to the 
identity map (see Figure~\ref{fig:Shor}C). 

It has been argued~\cite{Cao2015} that the three-bit precision of Monz 
\textit{et al} is insufficient since $2n$ bits are required for the algorithm 
to overcome Fourier leakage in general, to succeed with a bounded error rate 
required for scalability. However, for $N=15$ there is no Fourier leakage 
because of the power-of-two periods, and this is clear when building the 
quantum gate array. Therefore, measuring more qubits will only add noise, not 
precision. The distribution is completely described at two bits of precision. 
On the other hand, this also means that the process of generating the circuitry 
solves the factoring problem, since it is enough to know at which point the 
identity emerges from exponentiation. Thus, the experiment \cite{Monz2016} is 
not so much about factoring 15 through the Shor algorithm, but more a 
verification that the quantum circuit used behaves as expected when not using a 
compiled circuit.

We now present an experimental realization of Shor's algorithm, factoring 15. Please note that the motivation for doing this is not to factor 15, but to compare our construction with the current state-of-the-art. We use basically the same algorithmic setup, but employ the semiclassical Fourier transform and not of qubit recycling, and the emerging identity
operations are of course all omitted. Our setup is similar to that of Monz \emph{et al.}, without the qubit recycling, see
Figure~\ref{fig:Shor}B. Similar to Monz \emph{et al} we use the multiplication operators from
Markov et al \cite{Markov2012}, but avoid precomputing their effect on the initial state, see Figure~\ref{fig:Shor}C. 

Our setup runs on quantum logical gates from an improved version of the QSL
framework \cite{Patent,Johansson2016} inspired by Spekkens'
model, where a qubit is simulated by two classical bits; one for the
computational basis and one for the phase. A source of $\ket0$ or $\ket1$
initializes the computational bit to the bit value and randomizes the phase bit.
A measurement returns the value of the computational bit (and randomizes the
phase bit). The operation of the gates can be found in
Figure~\ref{fig:QSL_gates}, where Hadamard and CNOT are as in
~\cite{Johansson2016}. The new classically controlled $R_2$-gate XORs the
computational-bit onto the phase-bit and then inverts the computational-bit, if
the control is 1. The Toffoli construction is much improved, and is used to
construct the Fredkin gate needed for Shor's algorithm. The physical 
implementation is in 2-complementary reversible pass-transistor
logic, specifically using transmission gates~\cite{Caroe2012}. These are
constructed with currently available semiconductor technology, and operated at
an ambient temperature around 300~K.

The output probability distributions are estimated from $10^6$ samples for each element
$a\in\{2,4,7,8,11,13\}$, see Figure~\ref{fig:distribution}.
We see that the distributions for $a=7,8,13$ are not uniform
as predicted by quantum theory, but to contrast with the square statistical
overlap (SSO)~\cite{Chiaverini2005a} used by Monz et al. as a fidelity measure,
we get $\{$0.9999(1), 0.9999(1), 0.933(3), 0.984(2), 0.9999(1), 0.984(2)$\}$ for
$a\in\{2,4,7,8,11,13\}$ respectively (statistical errors as one standard 
deviation). Notably, the implementation gives the same probability (0.5) of returning a good candidate for $r$, as the ideal quantum subroutine.

In conclusion, we have created and implemented a framework consisting of
sources, gate array transformations, and measurements that can be used to run a 
quantum algorithm to a precision at least as good as state-of-the-art 
experiments factoring 15. The framework uses only classical resources, and the 
overhead is constant, in terms of memory and gate count (space and time). As 
any other physical implementation of quantum gates, our 
simulation suffers from systematic errors. This will result in an error 
propagation that suppresses the amount of useful information that we can 
retrieve when scaling the algorithm to larger numbers --- even though there is 
no practical restriction for doing so. Further work is needed to reduce these systematic errors, perhaps by altering the framework or using error 
correcting techniques, so that the framework becomes 
useful for larger instances. 

Most importantly, the QSL framework provides a fair comparison between quantum 
and classical computation. Even though it uses classical resources, it contains 
quantum-like degrees of freedom, and reproduces many phenomena of quantum 
mechanics including interference. This is enough to approach the general 
behavior of quantum algorithms. The remaining systematic errors point to 
properties of quantum gates and systems that the QSL framework 
\emph{does not} reproduce. Knowledge of this substantially narrows down the 
search area for what the truly quantum resources are, and also points to the 
properties that are necessary for providing a quantum advantage.
This is crucial information for present and future projects aiming at 
building a quantum computer.


\section*{References}
\printbibliography[heading=subbibliography, title={\null}]

\end{document}